\documentclass[reprint,NumberedRefs]{JASA}
\usepackage{amsmath,amsfonts,amssymb,amsthm,bm}

\usepackage{graphicx}

\renewcommand\vec{\mathbf}
\begin{document}
\title{Scattering from layered seafloors: Comparisons between theory and integral equations}
\author{Derek R. Olson}
\affiliation{Naval Postgraduate School, Monterey, CA 93943}
\email{dolson@nps.edu}
\thanks{Corresponding Author}

\author{Darrell Jackson}
\affiliation{Applied Physics Laboratory,  University of Washington, Seattle, WA, 98105 USA}

\date{\today}
\begin{abstract}
Acoustic scattering from layered seafloors exhibits dependence on both the mean geoacoustic layering, as well as the roughness properties of each layer. Several theoretical treatments of this environment exist, including the small roughness perturbation approximation, the Kirchhoff approximation, and three different versions of the small slope approximation. All of these models give different results for the scattering cross section and coherent reflection coefficient, and there is currently no way to distinguish which model is the most correct. In this work, an integral equation for scattering from a layered seafloor with rough interfaces is presented, and compared with small roughness perturbation method, and two of the small slope approximations. It is found that the most recent small slope approximation by Jackson and Olson is the most accurate when the root mean square (rms) roughness is large, and some models are in close agreement with each other when the rms roughness is small.
\end{abstract}

\maketitle

\section{Introduction}
The ocean floor contains variations in both its roughness and layering structure. At low frequencies, sound can interact with sub-bottom layers but the effects of roughness may be relatively small (at least for modest roughness). At high frequencies, the attenuation in the ocean bottom is higher, and the interaction with sub-bottom layers is reduced, but the effect of scattering may be more important. At intermediate frequencies, acoustic waves interact with both the sub-bottom layering, and roughness. It is often of practical interest to remotely sense properties of both sub-bottom layers, and the rough interfaces that separate them. Several models have been previously developed to solve the forward problem using a point source\cite{Tang2017,Olson2020,Pinson2016,Pinson2017,Holland2017}.

Models used for these purposes are limited in their applicability. The Kirchhoff approximation (KA) is restricted to angles close to the specular direction, and the small roughness perturbation method (SPM or perturbation theory) performs best away from specular. The small-slope approximation was introduced by Voronovich\cite{Voronovich1985}, and is applicable to the entire angular range for certain parameters of the rough interface\cite{Broschat1997}. Its original incarnation was for Dirichlet boundary conditions, but it has been applied to fluid, elastic \cite{Yang1994,Gragg2001}, and poroelastic\cite{Yang2002} halfspaces.

Recently, the small-slope approximation has been expanded to encompass layered media, but there are three competing models. One  small-slope approximation for layered media was developed by Jackson\cite{Jackson2013}, but is not explored here due to its strange behavior for slow sediment layers. Another small-slope approximation was developed by Gragg and Wurmser\cite{Gragg2005} in the acoustics literature, and later by Berrouk \textit{et al}.\cite{Berrouk2014} in the electromagnetics literature. It is denoted SSL2 in this work. The last, and most complicated small slope approximation was developed by Jackson and Olson\cite{Jackson2020}, and is denoted SSL3. The SSL$n$ convention comes from Jackson and Olson\cite{Jackson2020}, and is retained here. As shown by Jackson and Olson\cite{Jackson2020}, all of these models disagree for certain roughness and layer geoacoustic properties. This ambiguity is troubling. Although SSL3 has the most physically relevant motivation, it is not clear which approximation should be used in a given situation.

In this work, we remedy this ambiguity by providing comparison between SSL2, SSL3, SPM and the exact solution using integral equations. Two geoacoustic environments with two sets of roughness parameters each are used. The integral equation method is based on Monte-Carlo averaging, so individual realizations must be produced. A recent application of the Kirchhoff approximation\cite{Pinson2016} treats the the seafloor layering more faithfully than previous work, but is specialized to the point-source, point-receiver geometry, not plane waves. Since formally-averaged quantities such as the scattering cross section, are not available, we make no comparisons to this model in this work. Comparison to these models is certainly a fruitful area for future work. We find that SSL3 provides the best match with exact results for the scattering strength and coherent reflection coefficient. We do not present a systematic study of the region of validity for these models, although that is also a productive area for future work.

In Section \ref{sec:geometryEnvironment} we present the geometry and environment. The basic concepts for the models used here are presented in Section \ref{sec:models}. The integral equation method is detailed in Section \ref{sec:integralEquations}. Comparisons are made to theory in Section \ref{sec:results}. Discussion and conclusions are presented in Section \ref{sec:discussion}.
%
%
%
%
\section{Geometry and Environment}
\label{sec:geometryEnvironment}
\begin{figure}
	\centering
	\includegraphics[width=3.375in]{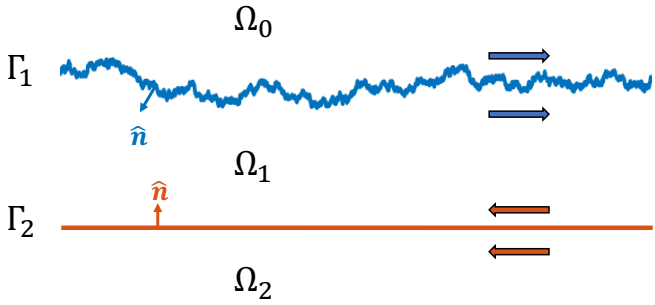}
	\caption{(color online) Layered environment and geometry. Although only the upper interface is depicted as rough, the integral equations defined here can be used with two rough boundaries. $\Gamma_n$ denotes each interface, and $\Omega_n$ denotes the medium immediately above $\Gamma_n$. The arrows show the direction of integration used in the integral equations developed in Sec.~\ref{sec:integralEquations}.}
	\label{fig:env}
\end{figure}

The geometry of the problem is shown in Fig.~\ref{fig:env}. Although arbitrary fluid layering is treated in theoretical work\cite{Jackson2020}, we limit the problem here to an overlying water column (a half-space), a fluid layer, and an underlying fluid half-space (which we refer to as the basement). These domains are denoted as $\Omega_0$, $\Omega_1$, and $\Omega_{2}$ respectively. Each domain, $\Omega_n$ is bounded by one of two boundaries. $\Gamma_{1}$ bounds $\Omega_0$ from $\Omega_{1}$ and is the water-sediment interface. $\Gamma_2$ bounds $\Omega_{1}$ from $\Omega_{2}$, and is the interface between the sediment layer and the sediment basement. The  boundary of a domain $\Omega_n$ is denoted $\partial \Omega_n$,  with $\partial\Omega_0 = \Gamma_1$ , $\partial\Omega_{1} = \Gamma_1 \cup \Gamma_2$, and $\partial \Omega_2 = \Gamma_2$. The normal vectors associated with each of these boundaries are shown in Fig.~\ref{fig:env}. Note that both normal vectors point into $\Omega_{1}$. This property is important for derivation of the boundary integral equations.

Each domain, $\Omega_{n}$, is characterized by a phase speed $c_n$, density $\rho_n$, and dimensionless loss parameter, $\delta_{n}$. The complex sound speed in each domain can be written as
\begin{align}
   \tilde{c}_n = \frac{c_n}{1 + i\delta_n}.
\end{align}
The wavenumber in each domain is related to the complex sound speed through $k_n = \frac{\omega}{\tilde{c}_n}$, where $\omega$ is the acoustic angular frequency with units of radians per second. Dimensionless ratios are defined as $a_{c1} = c_{1}/c_0$, and $a_{c2} = c_{2}/c_{0}$ for sound speed, and $a_{\rho 1} = \rho_{1}/\rho_{0}$ and $a_{\rho 2} = \rho_{2}/\rho_0$.

The incident acoustic wave vector is specified by
\begin{align}
	\vec{k}_i = k_0 \hat{\vec{k}}_i,
\end{align}
where $\hat{\vec{k}}_i$ is the incident acoustic unit wave vector, and is given by
\begin{align}
	\hat{\vec{k}}_i = -\cos\theta_i \hat{\vec{x}} - \sin \theta_i \hat{\vec{z}},
\end{align}
where $\hat{\vec{x}}$ is the unit vector in the $x$ (horizontal) direction, and $\hat{\vec{z}}$ is the unit vector in the $z$ (vertical) direction. The scattered wave vector into $\Omega_0$, back into the water column, is similarly given by 
\begin{align}
	\vec{k}_s = k_0 \hat{\vec{k}}_s,
\end{align}
where $\hat{\vec{k}}_s$ is the scattered acoustic unit wave vector, and is given by
\begin{align}
	\hat{\vec{k}}_s = \cos\theta_s \hat{\vec{x}} + \sin \theta_s \hat{\vec{z}}.
\end{align}
The incident grazing angle $\theta_i$, and scattered grazing angle $\theta_s$, are both measured from the horizontal axis.

The rough interfaces are described in terms of their power spectra. Let $W_n$ be the power spectrum of the rough interface constituting $\Gamma_n$. The rough interface $\Gamma_n$ is specified by the function $f_n(x_n)$. The Fourier transform of $f_n(x)$ is denoted $F_n (k_{x})$ with wavenumber argument $k_x$. The power spectrum is defined by $W_n(k_{x1})\delta(k_{x2} - k_{x1} )=\langle F_n(k_{x1}) F_n(k_{x2})^\ast \rangle$, where the angle brackets denote ensemble averaging. The truncated power law roughness spectrum known as the ``von K{\'a}rm{\'a}n'' spectrum is used here, and is specified by
\begin{align}
	W_n(K_x) = \frac{w_{1n}}{\left(K_{0n}^2 + k_x^2\right)^{\gamma_{1n}/2}}
\end{align}
where $w_{1n}$ is the one-dimensional (1D) spectral strength for interface $n$ with units of m$^{3-\gamma_{1n}}$, $\gamma_{1n}$ is the dimensionless 1D spectral exponent, and $K_{0n}$ is the spectral cutoff for interface $n$ with units of rad/m. The mean square height for interface $n$ is denoted $h_n^2$, and is equal to the integral of $W_n$ over the real line. For the von K{\'a}rm{\'a}n spectrum, 
\begin{align}
	h_{1n}^2 = \frac{w_{1n}\sqrt{\pi}\Gamma\left( (\gamma_{1n}-1)/2 \right)}{K_{0n}^{\gamma_{1n}-1}\Gamma\left( \gamma_{1n}/2 \right)}.
\end{align}
Values of $\gamma_{1n}$ greater than unity result in a finite $h_{1n}^2$. The 1D correlation function for the $n$-th interface, $C_n(x)$ is defined as
\begin{align}
	C_n(x) &=h_{1n}^{-2} \langle f_n(x^\prime)f_n(x + x^\prime)\rangle .
\end{align}
For the von Karman spectrum, the correlation function is
\begin{align}
	C_n(x) &= \frac{2^{1 - \nu_n} }{\Gamma(\nu_n)}\left(K_{0n} x\right)^{\nu_n} K_{\nu_n}(K_{0n} x)
\end{align}
where $\nu_n=(\gamma_{1n} - 1)/2$. $K_\nu(x)$ is the modified Bessel function of the second kind with argument $x$ and order $\nu$. 
We assume the rough interfaces do not intersect. 
%
%
%
\section{Models for scattering from one-dimensional roughness}
\label{sec:models}
The three scattering models compared in this work are the SPM, SSL2 and SSL3. Since a complete description of these models is quite lengthy, only the elements will be provided here. The reader is referred to Jackson and Olson\cite{Jackson2020} where these models are presented in complete form. We focus on two quantities, the coherent reflection coefficient, and the scattering cross section, both of which are defined in terms of the T-matrix, $T(k_{sx},k_{ix})$. The T-matrix is a transfer function between an incident plane wave with horizontal wave vector $k_{ix}$, and a scattered plane wave with horizontal wave vector $k_{sx}$. The scattering cross-section due to 1D roughness, $\sigma(k_{sx},k_{ix})$,is defined as\cite{thorsos_jackson_1989}
\begin{align}
	\sigma(k_{sx},k_{ix}) &= \frac{k_{sz}^2}{k_0} C(k_{sx},k_{ix}),
\end{align}
\begin{align}	
	\begin{split}
	C(k_{sx},k_{ix}) \delta\left(k_{ix}- k_{ix}^\prime\right) &= \langle T(k_{sx},k_{ix})T^\ast(k_{sx},k_{ix}^\prime)\rangle\\
	& - \langle T(k_{sx},k_{ix})\rangle \langle T^\ast(k_{sx},k_{ix}^\prime)\rangle,
	\end{split}
\end{align}
where $C(k_{sx},k_{ix})$ is the incoherent second-moment of the T-matrix. The coherent reflection coefficient, $|R(k_{ix})|$ is defined as 
\begin{align}
	|R(k_{ix})|\delta\left(k_{sx} - k_{ix} \right) = \left\vert \langle T(k_{sx},k_{ix})\rangle \right\vert
\end{align}
where the delta function must be included on the left hand side since it is always present in the average T-matrix for stationary roughness. The quantity $|R|$ is actually the magnitude of the complex coherent reflection coefficient, $R$. However, for brevity, we refer to $|R|$ as the coherent reflection coefficient. The above definitions assume stationary roughness, and we will further assume Gaussian statistics for this random process. The coherent reflection coefficient is frequently used with an angular argument, $|R(\theta_i)|$, instead of the horizontal component of the wave vector.

For the scattering cross section, all models here use the factor $A_n(k_{sx},k_{ix})$ for the $n$-th interface, which is defined as\cite{Jackson2020}
\begin{align}
	\begin{split}
	A_n({k}_{sx},~{k}_{ix}) = &\frac{1}{a_{c (n-1)}^2 a_{\rho (n-1)}} \\
	                      \times & A_{n-1}(k_{sx}) A_{n-1}(k_{ix}){\tilde A}_n({k}_{sx},~{k}_{ix})~.
     \end{split}
	\label{eq:A_n}
\end{align}
where $ A_{n-1}(k_{x})$ is the amplitude of the downgoing plane wave coefficient in medium $n-1$ (just above interface $n$) due to a plane wave incident from medium $\Omega_0$, and 
\begin{align}
	\begin{split}
	{\tilde A}_n({k}_{sx},~{k}_{ix}) = &\frac{1}{2}\left\{ a_n[1 + V_n(k_{ix})] [1 + V_n(k_{sx})]\right. \\
 & \left.- b_n[1 - V_n(k_{ix})] [1 - V_n(k_{sx})]\right \}~,
 \end{split}
\end{align}
where
\begin{align}%
\label{eq:an_factor}%
a_n &= \left(1- \frac {a_{\rho (n-1)}} {a_{\rho n}}\right)\frac {k_{sx} k_{ix}}{ k_{n-1}^2}- 1 + 
\frac {a_{c (n-1)}^2 a_{\rho (n-1)}} {a_{c n}^2 a_{\rho n}} \\
b_n &= \left(\frac{a_{\rho n}}{a_{\rho (n-1)}} - 1\right) \beta_{n-1}(k_{ix}) \beta_{n-1}(k_{sx})~.
\end{align}
$V_n(k_{ix})$ is the flat-interface reflection coefficient of interface $\Gamma_n$ assuming an overlying infinite halfspace in medium $\Omega_{n-1}$. The sine of the angle in $\Omega_n$ is $\beta_n(k_{x})=\sqrt{1 - k_{x}^2/k_n^2}$. For the upper interface, $V_1(k_x)$ is
\begin{align}
	V_1(k_x) = \frac{V_1^H(k_x) + V_2^H(k_x)\mathrm{e}^{2 \mathrm{i} k_1 \beta_1(k_x) D} }{1+ V_1^H(k_x)V_2^H(k_x)\mathrm{e}^{2 \mathrm{i} k_1 \beta_1(k_x) D}}
	\label{eq:planeWaveModel}
\end{align}
where $D$ is the mean thickness of $\Omega_1$, and $V_n^H(k_x)$ is the reflection coefficient of the $n$-th layer assuming both sides consist of halfspaces - defined as
\begin{align}
	V_n^H(k_x) &= \frac{Z_n-1}{Z_n+1}\\
	Z_n &= \frac{a_{\rho n} a_{cn} \beta_{n-1}(k_x) }{a_{\rho (n-1)} a_{c(n-1)} \beta_{n}(k_x)}.
\end{align}

For perturbation theory, the 2D T-matrix for interface $n$ is
\begin{equation}
\label{T_n}
T_n^{\rm SPM}({k}_{sx},~k_{ix}) = \frac{\mathrm{i} k_0}{\beta_0(k_{sx})} 	A_n(k_{sx},~k_{ix})F_n(k_{sx}-k_{ix} )~.
\end{equation}
For SSL2, the 2D T-matrix for the layered, rough seafloor is.
\begin{align}
	\begin{split}
	T^{\rm SSL2}(k_{sx},~k_{ix}) &= -\frac{k_0}{ 2\pi \beta_0(k_{sx}) \Delta k_z} \\
		\times \sum_{n=1}^{N}A_n(k_{sx},~k_{ix}) & \int \mathrm{e}^{-\mathrm{i}(k_{sx} - k_{ix})x - \mathrm{i} \Delta k_z f_n(x)}\,\mathrm{d}x~.
		\label{eq:small_slope_SSL2}
	\end{split}
\end{align}
where $\Delta k_z = k_{sz} - k_{iz}$ is the difference between the vertical component of the scattered and incident wavenumbers.

SSL3 requires a version of $A_n(k_{sx},k_{ix})$ where interface $n$ has been displaced by an amount $f_n$, which is denoted $A_n(k_{sx},k_{ix},f_n)$. This expression is rather complicated, and the full version is presented in Eqs. (37) and (82) of Jackson and Olson\cite{Jackson2020}. The SSL3 T-matrix for interface $n$ in 2D geometry is
\begin{align}
	\begin{split}
	T_n^{\rm SSL3}(k_{sx},~k_{ix}) &= \frac{\mathrm{i} k_0}{2\pi  \beta_0(k_{sx}) }  \int \mathrm{e}^{-\mathrm{i}(k_{sx} - k_{ix}) x} \\
	\times &\int_0^{f_n(x)} A_n(k_{sx},~k_{ix},~f)\,\mathrm{d}f\,\mathrm{d}x~.
	\end{split}
	\label{eq:small_slope_SSL3_k_space}
\end{align}
The main difference between SSL2 and SSL3 is that in SSL3 the factor $A_n(k_{sx},k_{si},f_n)$ depends on the height of layer $n$, and the integral over space includes variations in the sediment layering due to roughness $f_n(x)$, whereas neither are true for SSL2. Because of this dependence on $f_n$, SSL3 takes into account changes due to roughness in the interference pattern produced by the layered seafloor, whereas SSL2 assumes that the interference pattern is unchanged by roughness. In this way, SSL2 may be thought of as a hybrid between SPM and a true small-slope approximation.

For the scattering cross section and coherent reflection coefficient, we refer the reader to Jackson and Olson\cite{Jackson2020}. Scattering strength for SPM is easy to compute using the formulas provided there. The coherent reflection coefficient for SSL2 is computed from Eq.~(74), and scattering strength from Eq.~(79) of that reference. For SSL3, the coherent reflection coefficient can be found in Eqs.~(83-87), and scattering strength in Eq.~(90-96), and (A1-A35). These formulae are omitted due to the large amount of space required to express these approximations and all of their definitions, and interpretations of the models will rely on expressions for the T-matrices presented above.

Jackson and Olson's analysis focused on two-dimensional roughness, whereas we consider one-dimensional (1D) rough interfaces in this work. These differences are minor for SPM, and formally averaging the SPM result is simple. The 1D version of SSL2 can be found by simply replacing the Kirchhoff integral, Eq.~(63) in Jackson and Olson\cite{Jackson2020} with
\begin{align}
	I_n^{1D}(\eta) = 2 \mathrm{e}^{-\eta^2 h_n^2} \int\limits_{0}^\infty \cos\left(\frac{\Delta K_x u}{k_0} \right)\left[ \mathrm{e}^{\eta^2h_n^2 C_n(u/k_0)} -1\right]\,\mathrm{d}u
\end{align}
where $\Delta K_x = k_{sx} - k_{ix}$ is the difference between the horizontal component of the the outgoing and incoming wave vectors. Similarly, the small slope integral, Eq.~(89) in Jackson and Olson\cite{Jackson2020} should be replaced by
\begin{align}
	I^{1D}_{ssln}(\eta_a,\eta_b) = \mathrm{e}^{-(1/2)(\eta_a - \eta_b)^2 h_n^2} I^{1D}_n(\eta_a \eta_b)
\end{align}
to compute SSL3.

\section{Integral Equations}
\label{sec:integralEquations}
Integral equations provide a method to produce the exact scattered pressure due to rough surfaces. Methods for pressure release\cite{Thorsos1988}, and fluid-fluid\cite{Thorsos2000} boundary conditions have been previously presented in the underwater acoustics literature. A numerical method for layered media was presented by Tang and Hefner\cite{Tang2012}, but its derivation was not based on an integral equation for that environment. Rather, in that reference, a discretized matrix equation is derived from the integral equation for a single interface, and a discretized matrix equation is given for the layered case via physical intuition. The method presented here is based on matching boundary conditions for three different integral equations. This method can be shown to be equivalent to the method of Tang and Hefner, after correcting a few errors on the diagonal terms, and rearranging the density ratio factors.

The pressure, $p_n$ in any domain $n$ must follow the Helmholtz equation in each domain,
\begin{align}
	\nabla^2 p_n + k_n^2 p_n = 0
\end{align}
where $\nabla^2$ is the Laplacian operator. The Green's function is the solution to the Helmholtz equation with a point source on the right-hand side. In two dimensions, the free space solution (i.e. without boundaries) in domain $n$ using a point source is
\begin{align}
	\nabla^2 G_n(\vec{r},\vec{r}_0) +k_n^2 G_n(\vec{r},\vec{r}_0) &= \delta(x-x_0)\delta(z-z_0)\\
	 G_n(\vec{r},\vec{r}_0) &= \frac{-\mathrm{i}}{4}H_0^{(1)}(k_n\left\vert \vec{r} - \vec{r}_0\right\vert)
\end{align}
where $\mathrm{i}=\sqrt{-1}$ is the imaginary unit, $\delta(x)$ is the Dirac delta function, and $H_0^{(1)}(x)$ is the Hankel function of the first kind of order zero, with argument $x$. The position vectors are defined as $\vec{r} = r_x \hat{\vec{x}} + r_z \hat{\vec{z}}$ and $\vec{r}_0 = r_{x0}\hat{\vec{x}} + r_{z0}\hat{\vec{z}}$. We denote the position vector restricted to $\Gamma_n$ as $\vec{r}_n$, and the normal vector as $\hat{n}_n$.


Within each domain, the pressure field satisfying a Helmholtz equation can be solved using the Helmholtz integral formula, also known as the Helmholtz-Kirchhoff Integral equation\cite{Pierce1994}. The pressure on the boundary can be expressed, in the absence of an incident pressure field, and with an outward-pointing normal vector (corresponding to the ``exterior'' boundary value problem), as
\begin{align}
	\alpha(\vec{r}_l) p_n(\vec{r}_l) = \mathbb{V}^n_{l,m}\frac{\partial p_n(\vec{r}_m) }{\partial \vec{n}_m}-\mathbb{K}^{n}_{l,m} p_n(\vec{r}_m)
	\label{eq:HKIE_oneDomain}
\end{align}
where the integral operators are defined as
\begin{align}
	\mathbb{V}^n_{l,m} [\phi(\vec{r}_l)] &= \int\limits_{\Gamma_m}G_n(\vec{r}_l,\vec{r}_m) \phi(\vec{r}_m)\,\mathrm{d}S_m \label{eq:defSLP}\\
	\mathbb{K}^n_{l,m} [\phi(\vec{r}_l)] &= \int\limits_{\Gamma_m}\frac{\partial G_n(\vec{r}_l,\vec{r}_m)}{\partial \vec{n}_m} \phi(\vec{r}_m)\,\mathrm{d}S_m \label{eq:defDLP}
\end{align}
$\phi(\vec{r})$ is an arbitrary square-integrable function, $dS_m$ indicates that the integration is carried out over the boundary with respect to the subscript variable, and $\partial/\partial \vec{n}_m = \hat{\vec{n}}_m \cdot \nabla_m$ is the normal derivative with respect to the $m$ argument (as opposed to $l$). The subscript of $l,m$ on the integral operators indicates that integration is carried out along $\Gamma_m$, and the operator output is a function defined on $\Gamma_l$. The parameter $\alpha$ is equal to $\beta/(2\pi)$, where $\beta$ is the angle subtended by the tangent lines on each side of a given point. For a smooth surface, $\alpha=1/2$ at all points. In this work, we form the integral equation along a piecewise continuous, non-smooth surface, and must calculate $\alpha$ at each point. The operator $\mathbb{V}$ is commonly referred to as the single-layer potential operator, and $\mathbb{K}$ as the double-layer potential operator. In this work, the exterior form of the Helmholtz integral equation is used with domains having a single boundary - only $\Omega_0$ and $\Omega_2$

We may also form the equivalent integral equation for a domain with inward-pointing normal vector (the ``interior'' boundary value problem). In this case, the interior integral equation is defined for $\Omega_1$ only, which is bounded by $\Gamma_1$ and $\Gamma_2$. Here we write the integral operators on each boundary separately, giving
\begin{align}
	\begin{split}
	\left(1 - \alpha(\vec{r}_1)\right) p_1(\vec{r}_1) &= -\mathbb{V}^1_{1,1}\frac{\partial p_1(\vec{r}_1)}{\partial n_1} + \mathbb{K}^{1}_{1,1} p_1(\vec{r}_1) \\
	&- \mathbb{V}^1_{1,2}\frac{\partial p_1(\vec{r}_2) }{\partial n_2}+\mathbb{K}^{1}_{1,2} p_2(\vec{r}_2)
	\end{split}
\\[2ex]
	\begin{split}
		\left(1 - \alpha(\vec{r}_2)\right) p_1(\vec{r}_2) &= -\mathbb{V}^1_{2,2}\frac{\partial p_1(\vec{r}_2)}{\partial n_2} + \mathbb{K}^{1}_{2,2} p_1(\vec{r}_2)\\
		& - \mathbb{V}^1_{2,1}\frac{\partial p_1(\vec{r}_2)}{\partial n_2}+ \mathbb{K}^{1}_{2,1} p(\vec{r}_1).
		\end{split}
	\label{eq:HKIE_oneDomain_interior}
\end{align}
although written separately, these equations should be thought of as being a single integral equation, since the first specifies the pressure on the upper boundary, and the second specifies the pressure on the lower boundary. Both are required for a solution of the boundary value problem.

To form an integral equation for the union of all domains, continuity conditions for pressure and normal velocity are enforced between all domains, keeping track of the normal vector direction. The incident pressure from domain $\Omega_0$ is added to the right hand side of the integral equation for that domain, and terms are rearranged to give the integral equations in terms of the following matrix of operators
\begin{widetext}
\begin{align}
	\left[
	\begin{array}{cccc}
		\alpha(\vec{r}_1)\mathbb{I} + \mathbb{K}^{0}_{1,1} & -\mathbb{V}^{0}_{1,1} & & \\
	\left(1 - \alpha(\vec{r}_1) \right)\mathbb{I} - \mathbb{K}^{1}_{1,1} & a_{\rho 1} \mathbb{V}^{1}_{1,1} & - \mathbb{K}^{1}_{1,2} & a_{\rho 1} \mathbb{V}^{1}_{1,2} \\
		- \mathbb{K}^{1}_{2,1} & a_{\rho 1} \mathbb{V}^{1}_{2,1} & \left(1 - \alpha(\vec{r}_2) \right)\mathbb{I}  - \mathbb{K}^{1}_{2,2} & a_{\rho 1} \mathbb{V}^{1}_{2,2} \\
			 & 	 & \alpha(\vec{r}_2)\mathbb{I} + \mathbb{K}^{2}_{2,2} & -a_{\rho 2} \mathbb{V}^{2}_{2,2} \\
	 \end{array}
	 \right]
	 \left[
	 \begin{array}{c}
		 p_0(\vec{r}_1) \\
		 \frac{\partial p_0(\vec{r}_1)}{\partial n_1} \\
		 p_1(\vec{r}_2) \\
		 a_{\rho1}^{-1}\frac{\partial p_1(\vec{r}_2)}{\partial n_2}
	 \end{array}
	 \right]
	 =
	 \left[
	 \begin{array}{c}
		 p_i(\vec{r}_1) \\
		 \\
		 \\
		 \\
	 \end{array}
	 \right],
	 \label{eq:multiLayerIntegralEquation}
 \end{align}
\end{widetext}
where a blank spot in a matrix denotes either a zero operator or a variable that is identically zero, and $\mathbb{I}$ denotes the identity operator (which maps a function onto itself). The right hand side of this system of integral equations is the incident pressure on $\Gamma_1$ from $\Omega_0$ in the first row, and is zero for all other rows. The unknown variables consist of the pressure in $\Omega_0$ and $\Omega_1$, as well as their normal derivatives. Note that the normal derivative for $\Omega_1$ has the factor $a_{\rho 1}^{-1}$, which is due to the boundary conditions for the continuity of the normal velocity across $\Gamma_2$.

In this equation, the direction of integration determines the direction of the unit normal vector. The convention followed here is that the normal vector points to the right of the integration direction along each boundary, $\Gamma_n$. In Fig.~\ref{fig:env}, the direction of integration along each boundary, and in each domain has been specified. In $\Omega_1$, which has both boundaries, the integration can be thought of being in the clockwise direction, to the right on the top, and to the left on the bottom. Formally, the integral should be closed in $\Omega_1$ between $\Gamma_1$ and $\Gamma_2$, but this part of the integral may be neglected if the pressure field decays to zero, which we assume here.

Extensions of this method to multiple layers can be made by formulating the Helmholtz-Kirchhoff integral equation (HKIE) in each domain, and matching boundary conditions. A systematic method to perform this type of calculation was presented by von Petersdorff and Leis\cite{vonPetersdorff1989}, although their analysis uses the operators in (\ref{eq:defSLP}) and (\ref{eq:defDLP}) and their normal derivatives (the adjoint double layer, and hypersingular potential operators respectively -- both of which are not used here). Although the method of von Petersdorff and Leis has superior stability and numerical conditioning than the method used here, it is more complicated due to the hypersingular operator, which is difficult to implement numerically. The numerical condition number for the method detailed in this work has been found to be adequate for our purposes (on the order of $10^6$ or $10^7$).

The incident field used here is an approximation to a plane wave developed by Thorsos\cite{Thorsos1988}. This field is incident  from $\Omega_0$ onto interface $\Gamma_1$, and takes the form (for our time convention)
\begin{align}
	p_i(\vec{r}_1,f) &= p_i \mathrm{e}^{\mathrm{i} \vec{k}_i\cdot \vec{r}_1 \left(1 + w(\vec{r}_1)\right) - \left(x_1 - z_1\cot\theta_i \right)^2/g^2} 	\label{eq:taperedPlaneWave}\\
		w(\textbf{r}_1) &= \frac{2\left( x_1 - z_1 \cot\theta_i \right)^2 / g^2 - 1 }{( k_0 g \sin\theta_i)^{2}} ,
  \label{eq:wCorrection}
\end{align}
where $g$ is a parameter controlling the width of the incident field, and $p_i$ is the complex pressure amplitude. The 3 dB angular width of this beam is
\begin{align}
	\Delta \theta =\frac{2\sqrt{2 \log(2)}}{k_0 g \sin\theta_i},
\end{align}
As the product $k_0 g$ grows large, the incident field better approximates a plane wave, and it is valid at lower grazing angles. The angular width increases as $\theta_i$ decreases, so small grazing angles are more computationally demanding for numerical solution of scattering problems\cite{Thorsos1988}.

These integral operators can be discretized using standard techniques, such as the boundary element method (BEM)\cite{Wu2000,Thorsos1988}. In this work, these operators were discretized using the collocation method with linear basis functions to approximate the pressure and normal derivative, resulting in a square matrix for each of the integral operators. The matrices were assembled into a fully discrete block matrix according to Eq.~(\ref{eq:multiLayerIntegralEquation}).
		
After the pressure and pressure normal derivative on each boundary is found, it is propagated to the far-field using the HKIE. If the pressure in $\Omega_0$ is sought, then this becomes
\begin{align}
	p_0(\vec{r}_f) = \mathbb{V}^0_{f,1}\frac{\partial p_0(\vec{r}_1)}{\partial n_1} - \mathbb{K}^{0}_{f,1} p_0(\vec{r}_1)
	\label{eq:HKIE_toField}
\end{align}
where the subscript $f$ denotes the field pressure point locations. The pressure in other domains can be found from the integral equations for that domain. Once the field pressure is found, the scattering cross section can be estimated by\cite{Thorsos1988}
\begin{align}
	\sigma = \frac{r}{L^\prime} \frac{\langle \left\vert p_0(\vec{r}_f)\right\vert^2 \rangle}{|p_i|^2 }
\end{align}
where $r$ is the distance from the center of the top mean interface to the field point, and
\begin{align}
	L^\prime= g \sqrt{\frac{\pi}{2}} \left[1 - \frac{0.5(1 + 2\cot^2\theta_i)}{(k g \sin\theta_i)^2} \right]
\end{align}
is the effective ensonified length of the rough interface. The angle brackets denote ensemble averaging.
 
The coherent reflection coefficient is a bit more difficult to estimate. Instead of an analytic formulation, we follow the method used by Thorsos\cite{Thorsos1990}. We compare the scattered pressure due to the rough layered environment to the scattered pressure in $\Omega_0$ due to a flat, rigid boundary of the same length, $p_{0flat}$. Namely,
\begin{align}
	|R_c| = \left\vert\frac{ \langle p_0(\vec{r}_f)\rangle  }{p_{0flat}(\vec{r}_f)}\right\vert.
\end{align}
These calculations use the same tapered incident field.

\section{Results}
\label{sec:results}
We present results for two different geoacoustic environments. The first environment has a layer with a greater sound speed than that of water, where $a_{c1}=1.05$, $a_{\rho 1}=1.8$, $a_{c2}=1.8$, $a_{\rho 2}=2.5$. The attenuation parameters are  $\delta_1=0.01$ and $\delta_{2}=0.02$. We call this environment the ``fast layer.'' The second environment is a slow mud layer overlying a fast basement with $a_{c1}=0.99$, $a_{\rho 1}=1.4$, $a_{c2}=1.8$, $a_{\rho 2}=2.5$. The attenuation parameters here are set to $\delta_1=0.0005$ and $\delta_{2}=0.02$, since softer sediments typically have smaller attenuation coefficient values. This environment is called the ``slow layer.'' Geoacoustic parameters are summarized in Table ~\ref{tab:geoacoustic_parameters}. These geoacoustic properties correspond to the second and third geoacoustic environments presented in Jackson and Olson\cite{Jackson2020}. The acoustic frequency was set to 2 kHz, with $\omega \approx 12.6\times10^{3}$ rad/s.

\begin{table}
\caption{Geoacoustic parameters used in numerical examples. All computations use water sound speed $c_0 = $1500 m/s, and density $\rho_0=1000$ kg/m$^3$, although only the ratios are important.}
\label{tab:geoacoustic_parameters}
\begin{tabular}{c c c c c c}
\hline \hline
Case & Domain & Thickness & Sound Speed & Density  & Loss  \\
     &           &  (m) & Ratio     & Ratio & Parameter \\   
\hline
Fast & 1 & 1 & 1.05 & 1.8 &  0.02  \\
Layer & 2 & $\infty$ & 1.8 & 2.5 &  0.01  \\
\hline
Slow & 1 & 1 & 0.99 & 1.4 &  0.005  \\
Layer& 2 & $\infty$ & 1.8 & 2.5 &  0.01  \\
\hline
\hline
\end{tabular}
\end{table} 

In all results, interface 2 is smooth and interface 1 is rough (except for the integral equation test case). Two sets of roughness parameters for each environment are used. One set has small $k_0 h_1$, and the other has larger $k_0 h_1$. These two parameter sets are presented in Table \ref{tab:roughness_parameters}. Note that $w_{11}$ and $\gamma_{11}$ for the large roughness case are the 1D equivalent to the parameters in the examples presented in Jackson and Olson \cite{Jackson2020}. Formulas from Appendix D and the errata list of Jackson and Richardson\cite{Jackson2007} were used to perform this conversion. In this table the rms height of each interface multiplied by $k_0$ is shown, as is the rms height divided by the average layer thickness, $D$, set to 1 m.

\begin{table}
	\caption{Roughness spectrum parameters used in numerical examples.}
	\label{tab:roughness_parameters}
\begin{tabular}{c  c  c  c   c c}
	\hline \hline
	Case & $w_{11}$  & $\gamma_{11}$ & $K_{01}$ & $k_0 h_1$ & $h_1/D$\\
	     & [m$^{3-\gamma_{11}}$] & - &    [rad/m]& -        & - \\
  	\hline 
	Large $k_0 h_1$ & 2$\times10^{-3}$ & 2 & 1 & 0.66 & 0.079\\
	Small $k_0 h_1$ & 2$\times10^{-4}$ & 2 & 1 & 0.21 & 0.025\\
	\hline \hline
\end{tabular}
\end{table}

The integral equation results used 48 independent roughness realizations. The incident field width parameter, $g$, was set to $40\lambda$, which limited the range of grazing angles over which the integral equation results are valid. At $25^\circ$ grazing, the incident field relative angular width is about 5\%, and is 10\% at 18$^\circ$ grazing. Therefore, conservatively, results should be trusted above 25 degrees, but plots are shown down to 18 degrees. The total surface length of the realizations was set to $L=5g$, so that multiple reflections between the interfaces could be captured accurately. This value was chosen by gradually decreasing the value of $L$ until noticeable effects were seen (starting at $L=16g$). The surface was sampled at $\Delta x = \lambda/16$, which is a rather small sampling interval, but was chosen because coarser sampling did not converge within 1 dB. The rough surfaces are generated using the spectral method of Thorsos\cite{Thorsos1988} with the specified sampling interval. However, the power spectrum was low-pass filtered so that slopes at very small scales did not cause numerical issues with the discretized integral equations. The power spectra at wavenumbers between $6k_0$ and $6.5k_0$ were smoothly tapered to zero using a raised cosine function (inspired by LePage and Schmidt\cite{LePage2003}), and were set to zero between $6.5k_0$ and $8k_0$ (the Nyquist wavenumber). This transition region corresponded to about 100 points of the sampled wavenunber domain, with a total of 3480 points.

With these parameters, SPM was the fastest model to compute, as it required only evaluating functions at the angular and wavenumber arguments. SSL2 was the next fastest, but required evaluating the Kirchhoff integral for each incident and scattered angle. The integration was performed using the trapezoidal rule, and the time required was 0.5 s for each scattering strength figure presented below. SSL3 required computing the Kirchhoff integral many times for each incident and scattered direction, and required 28 minutes for each scattering strength figure presented below. The integral equation calculations for these parameters took about 12 hours for all 48 realizations. Times listed here were for an equivalent single-core processor, but we used a quad-core processor at 4.2 GHz and parallel for-loops to speed up the calculations.

\subsection{Integral Equation Validation}
\begin{figure}
  \centering
    \includegraphics[width=3.375in]{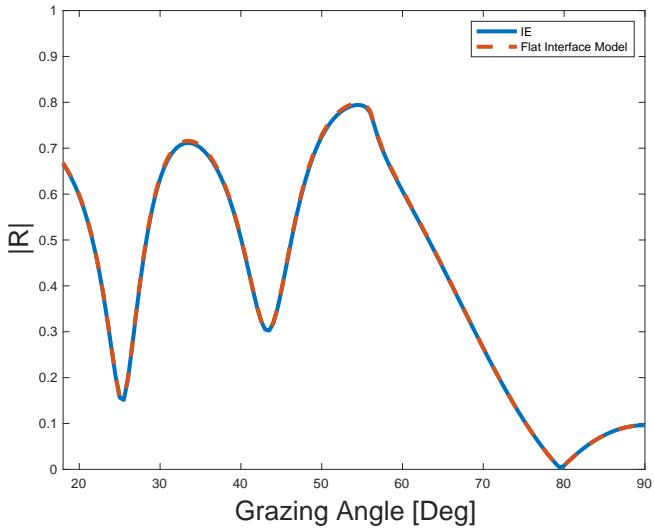}
  \caption{(color online) Flat-interface integral equation result compared with the plane wave reflection coefficient model.}
  \label{fig:highContrast_IE_vs_flat}
\end{figure}

The first result is a validation of the integral equation. The fast layer geoacoustic properties are used with both interfaces flat. The numerically calculated reflection coefficient is compared to the theoretical plane wave reflection coefficient, Eq.~(\ref{eq:planeWaveModel}). These two curves are compared in Fig.~\ref{fig:highContrast_IE_vs_flat}. The lower limit of the grazing angles is set to 18 degrees, since the angular width of the incident field and finite length of the surfaces that enter into the integral equations can cause discrepancies at low grazing angles. The model-IE comparison is quite good, although the IE result is slightly less than the model at small grazing angles, by about 1\%. 

\subsection{Fast layer}
\begin{figure}
	\centering
	\includegraphics[width=3.375in]{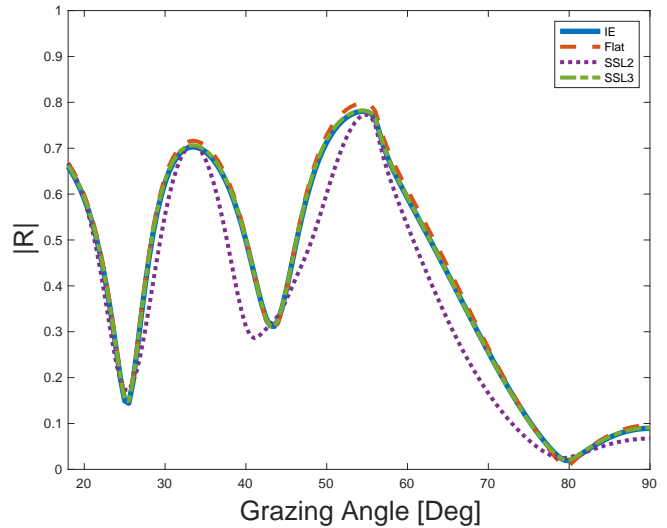}
	\caption{(color online) Coherent reflection coefficient for fast layer, small roughness case.}
	\label{fig:coh_refl_fast_small}
\end{figure}

A comparison for the coherent reflection coefficient between the theoretical models and integral equations for the fast layer with small roughness is presented in Fig.~\ref{fig:coh_refl_fast_small}. The small rms roughness causes the integral equation coherent reflection coefficient to depart only slightly from the flat-interface cases. SSL3 agrees with the integral equation result in this case, but SSL2 shows noticeable departures from all other models and integral equations. This figure serves to show that for small values of the rms roughness, defined here as $k_0 h_1 \approx0.2$, SSL3 is accurate for the coherent reflection coefficient, but SSL2 is less accurate, although not by much.

\begin{figure}
	\centering
	\includegraphics[width=3.375in]{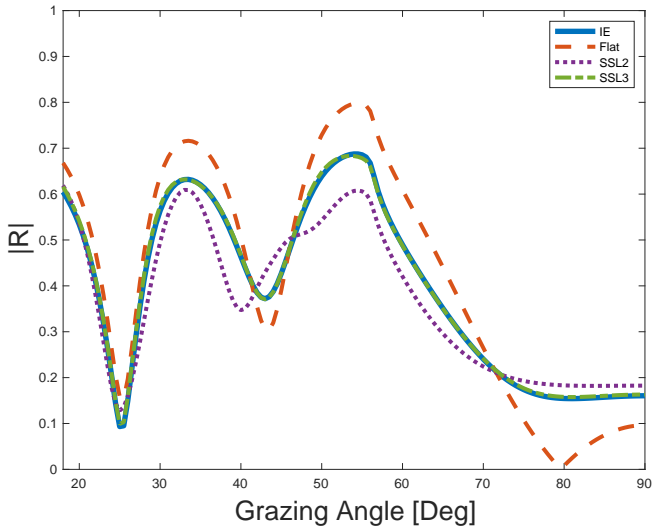}
	  \caption{(color online) Coherent reflection coefficient for fast layer, large-roughness case.}
	  \label{fig:coh_refl_fast_large}
\end{figure}

Results for the coherent reflection coefficient for large roughness are presented in Figure~\ref{fig:coh_refl_fast_large}. The integral equation departs significantly from the flat interface case, especially near 80 degrees grazing angle, and near the peaks. SSL3 is the best model presented here, although it has some small errors near the peaks. SSL2 follows the integral equation less closely. A notable difference is that SSL2 has a different local minimum near 45 degrees grazing angle than both the integral equation, flat interface, and SSL3. We may conclude that when the roughness is increased, SSL3 is a more physically realistic model, since it matches the integral equation results. Physically, this improved accuracy is due to the fact that SSL3 accounts for changes to the interference pattern when roughness is present in a layered seafloor.

\begin{figure}
	\centering
	\includegraphics[width=3.375in]{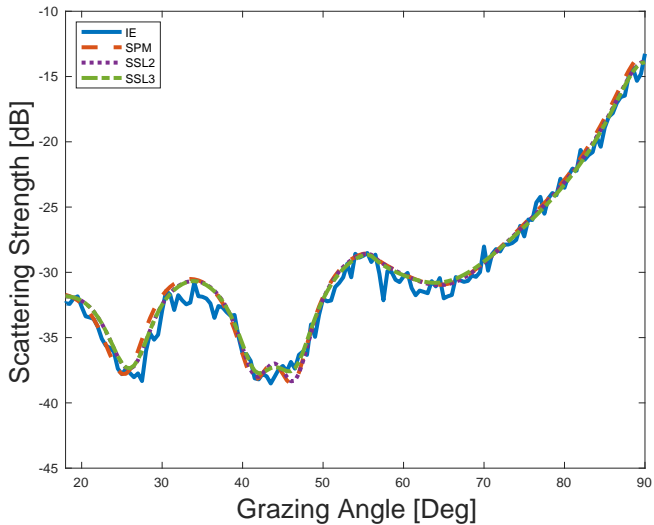}
	\caption{(color online) Scattering strength results for the fast layer, small roughness case.}
	\label{fig:scatt_strength_fast_small}
\end{figure}

Scattering strength results from the fast layer case with small rms roughness are presented in Fig.~\ref{fig:scatt_strength_fast_small}. The integral equation result is shown, along with SPM, and both SSL2 and SSL3. For these values of the geoacoustic and roughness parameters, all three models agree quite well with each other, and the models appear to fall within the uncertainty of the Monte-Carlo simulations. For these values of the roughness and geoacoustic parameters, we conclude that all models examined perform adequately for the scattering cross section. This figure gives confidence that SSL2 and SSL3 agree for small rms roughness, which they should in the limit as $k_0 h_1\to0$. The agreement of all the models with the integral equation results gives confidence that the implementations of both the theoretical models and integral equations are sound. It is interesting to note that for these roughness and geoacoustic parameters, SSL2 and SSL3 agree for scattering strength, but not for the coherent reflection coefficient.

\begin{figure}
	\centering
	\includegraphics[width=3.375in]{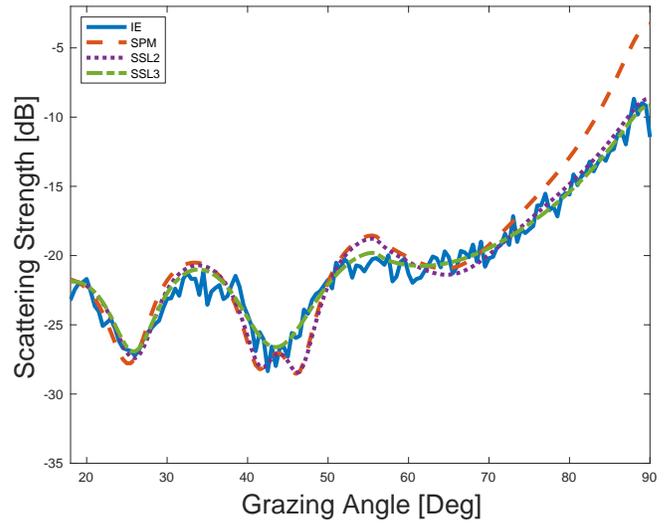}
	\caption{(color online) Backscattering strength comparison for fast layer, large roughness case.}
	\label{fig:scatt_strength_fast_large}
\end{figure}

When the roughness is increased in Fig.~\ref{fig:scatt_strength_fast_large}, all the models depart from one another, but only slightly. Perturbation theory becomes inaccurate near the specular direction, which is expected, but also contains some small errors at moderate angles. SSL2 performs better than perturbation theory near the specular direction, but still contains moderate errors compared to the integral equation at moderate angles. SSL3 performs better than SSL2, notably near 55 degrees grazing. Another notable difference between SSL2 and SSL3 are the changes in shape near 43 degrees grazing. Close examination of this region shows that SSL3 has a much different shape than both SSL2 and SPM. It appears that SSL3 is predicting an alteration of the interference pattern compared to SPM. The uncertainty of the integral equation results is too large to make a determination about which model is correct, but it appears that SSL3 follows the IE curve more closely between about 45-50 degrees grazing.

\subsection{Slow Layer}
\begin{figure}
  \centering
    \includegraphics[width=3.375in]{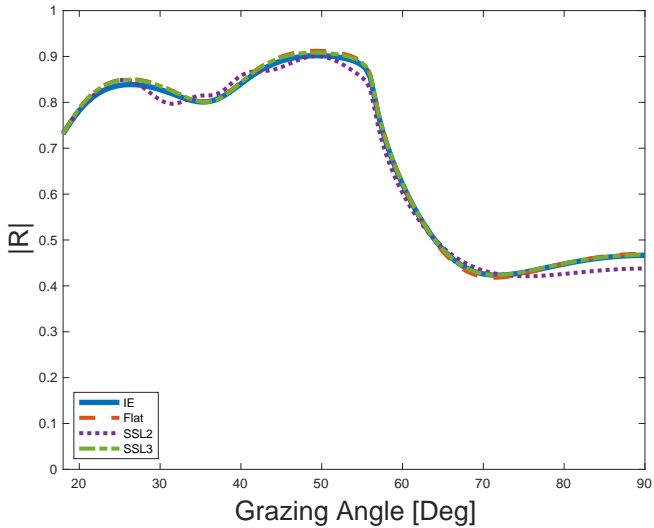}
  \caption{(color online) Coherent reflection coefficient for slow layer, small roughness case.}
  \label{fig:coh_refl_slow_small}
\end{figure}

The coherent reflection coefficient for the slow layer geoacoustic environment with small roughness properties is shown in Fig.~\ref{fig:coh_refl_slow_small}. It is compared with SPM, SSL2, and SSL3.  The oscillations in the coefficient are small compared to that of the fast layer. For this case, the integral equation, flat interface, and SSL3 agree quite well each other. SSL2 contains small discrepancies compared to the integral equation, although less than the fast layer case with small roughness. All the models presented here perform an adequate job for this small roughness case -- even the assumption of a flat interface.

\begin{figure}
	\centering
	\includegraphics[width=3.375in]{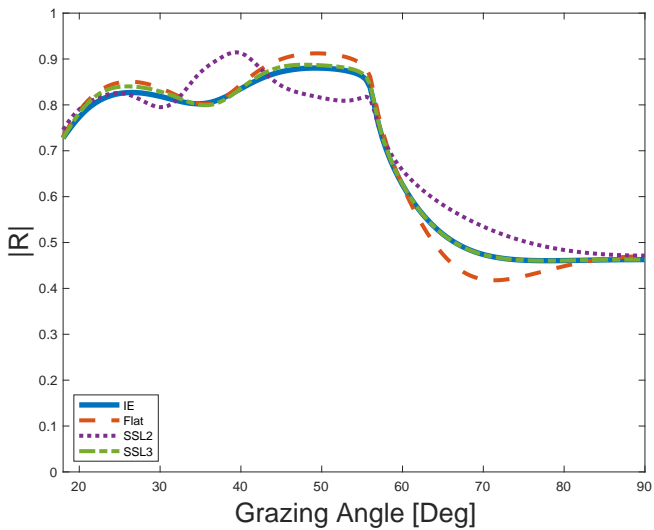}
	\caption{(color online) Coherent reflection coefficient for slow layer, large roughness case.}
	\label{fig:coh_refl_slow_large}
\end{figure}

The coherent reflection coefficient for the slow layer with large roughness is plotted in Fig.~\ref{fig:coh_refl_slow_large}. SSL3 is the best model here with an error of less than 1\%. SSL2 departs significantly from the integral equation result, especially near 60-70 degrees grazing and around 35 degrees grazing. SSL2 and the flat-interface model all have error of less than 10\%. Based on these small errors we conclude that the roughness present for this case has a small effect on the reflection coefficient for a slow layer, but is best modeled by SSL3. However, this conclusion cannot be extended to the scattered field in general, as will be seen in the next example.

\begin{figure}
  \centering
    \includegraphics[width=3.375in]{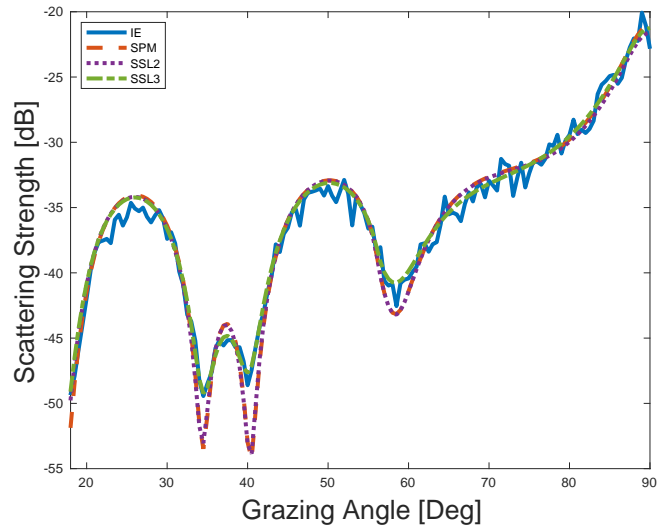}
  \caption{(color online) Backscattering strength comparison for slow layer, small roughness case}
  \label{fig:scatt_strength_slow_small}
\end{figure}

Scattering strength from the slow layer with small roughness is presented in Fig.~\ref{fig:scatt_strength_slow_small}, and the IE results are compared to the same models. The shape of the IE curve is much different from the fast layer case, and there are some deep nulls at several angles. SSL3 follows the IE curve the best. SSL2 and SPM agree with the integral equation result over most of the angular range, except for the three local minima present, near 35, 40, and 60 degrees grazing. Here, SPM and SSL2 underestimate the scattering cross section. SSL3 provides the correct fit near these grazing angles. If the attenuation of $\Omega_1$ were increased slightly, then SSL2 and SPM would match the IE result nearly as well as SSL3. 

\begin{figure}
	  \centering
   \includegraphics[width=3.375in]{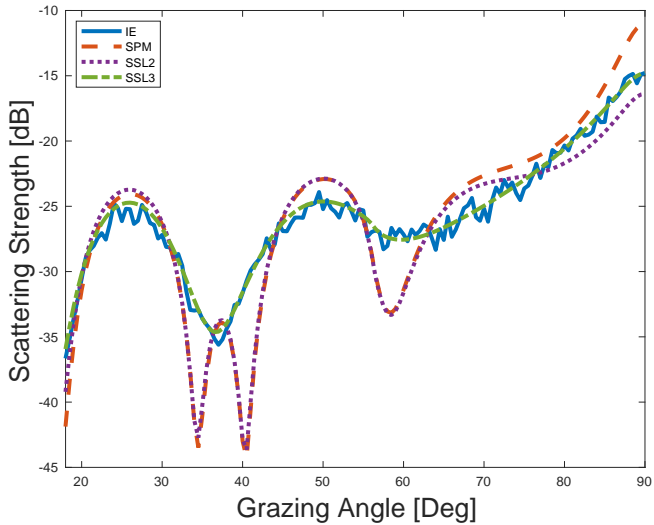}
  \caption{(color online) Backscattering strength comparison for slow layer, large roughness case}
  \label{fig:scatt_strength_slow_large}
\end{figure}

Scattering strength results from the slow layer case with larger rms roughness are presented in Fig.~\ref{fig:scatt_strength_slow_large}. The integral equation result is shown, along with SPM, and both SSL2 and SSL3. For these values of the geoacoustic and roughness parameters, SSL2 and SPM agree over all but the largest grazing angles. SSL3 disagrees with both SSL2 and SPM close to specular, and near the local minima away from the specular direction. The integral equation result agrees quite well with SSL3, but not SSL2 or SPM. The large differences between SSL2 and SSL3 are surprising, given the similar results presented for the coherent reflection coefficient in Fig.~\ref{fig:coh_refl_slow_large}. However, we note that the scattering cross section depends on the factors $(1 \pm V_1(\theta_i))$ and $(1\pm V_1(\theta_s))$, as seen in Eq.~(\ref{eq:A_n}). Since the magnitude of $V_1(\theta)$ is close to unity for the slow layer, small changes in $V_1(\theta)$ can lead to large relative changes in $(1\pm V_1(\theta))$, depending on the sign of $V_1(\theta)$. We can conclude from this example that for a slow layer, increasing the roughness causes large discrepancies between SSL2 and SSL3, and that SSL3 is the most accurate model for these paratmeters, to within about 0.3 dB.

\section{Discussion and Conclusion}
\label{sec:discussion}
In Section~\ref{sec:results}, we have seen that for the larger values of spectral strength, SSL3 departs from SSL2, which was shown previously in Jackson and Olson\cite{Jackson2020}. We have also seen that the Monte-Carlo integral equation method agrees very well with SSL3 in the large-roughness cases, and agrees with all the models in the small roughness case. In Jackson and Olson\cite{Jackson2020}, it was postulated that the disagreement between SSL2 and SSL3 was due to the fact that SSL3 accounts for changes to the interference pattern due to changes in layer thickness caused by the rough interfaces. The agreement between SSL3 and the integral equation supports the conclusion that this effect is indeed present in scattering from layered surfaces. It remains to be seen whether these differences can be seen in field experiments.

To investigate what is causing these large differences, another numerical experiment was performed with the same sound-speed and density as the fast layer, but with the attenuation of the slow layer, and the large roughness parameters. These results are not shown, but the decreased attenuation of the fast layer showed deep nulls in SSL2 and SPM that were not apparent in SSL3 or the IE results. When large roughness is present, its effect is more pronounced if the attenuation is small, for both a fast and a slow layer. We can conclude that the presence of roughness changes the interference pattern caused by the layering structure, and this effect is more pronounced with small values of the attenuation coefficient in $\Omega_1$.

The confirmation of the differences between SSL3 and SPM has implications for geoacoustic inversion of layered rough seafloors. The alteration of the interference pattern due to roughness could cause inversion schemes that use the reflection coefficient\cite{Dettmer2010,Holland2012}, or scattering strength\cite{Steininger2013}, to provide incorrect results if an inappropriate model is used, such as the flat-interface assumption for the reflection coefficient, or the SPM for the scattering cross section. SSL3 provides a promising model to use in such cases.


This work was focused on deciding between competing models for layered rough interfaces for a few sets of geoacoustic and roughness parameters. A systematic study of the validity of each of these scattering models was not performed, but would be a valuable avenue for future work. The integral equation methods presented here could be used for such a study.

\section*{Acknowledgments}
Funding was provided by the US Office of Naval Research.


\end{document}